\documentclass[epsfig,graphics,graphicx,eqsecnum,showpacs,showkeys,aps]{revtex4}
\usepackage{graphicx}
\usepackage{colordvi}

\begin{document}

\title{Canonical transformation for stiff matter models in quantum cosmology}

\author{C. Neves\footnote{E-mail: cliffordneves@uerj.br}}

\author{G. A. Monerat\footnote{E-mail: monerat@uerj.br}}

\affiliation{Departamento de Matem\'{a}tica, F\'{\i}sica e Computa\c{c}\~{a}o, \\
Faculdade de Tecnologia, \\ 
Universidade do Estado do Rio de Janeiro,\\
Rodovia Presidente Dutra, Km 298, P\'{o}lo
Industrial,\\
CEP 27537-000, Resende-RJ, Brazil.}

\author{G. Oliveira-Neto\footnote{E-mail: gilneto@fisica.ufjf.br}}

\affiliation{Departamento de Fisica, \\
Instituto de Ci\^{e}ncias Exatas, \\ 
Universidade Federal de Juiz de Fora,\\
CEP 36036-330 - Juiz de Fora, MG, Brazil.}

\author{E. V. Corr\^{e}a Silva\footnote{E-mail: evasquez@uerj.br}}

\author{L. G. Ferreira Filho\footnote{E-mail: gonzaga@fat.uerj.br}}

\affiliation{Departamento de Matem\'{a}tica, F\'{\i}sica e Computa\c{c}\~{a}o, \\
Faculdade de Tecnologia, \\ 
Universidade do Estado do Rio de Janeiro,\\
Rodovia Presidente Dutra, Km 298, P\'{o}lo
Industrial,\\
CEP 27537-000, Resende-RJ, Brazil.}

\date{\today}

\begin{abstract}
In the present work we consider Friedmann-Robertson-Walker 
models in the presence of a stiff matter perfect fluid and a cosmological
constant. We write the superhamiltonian of these models using the
Schutz's variational formalism. We notice that the resulting superhamiltonians
have terms that will lead to factor ordering ambiguities when they are written as operators. 
In order to remove these ambiguities, we introduce appropriate coordinate 
transformations and prove that these transformations are canonical using the 
symplectic method.
\end{abstract}



\maketitle

Recently, many authors have used the Schutz's variational formalism in order to 
write the superhamiltonian of Friedmann-Robertson-Walker (FRW) quantum cosmological 
models coupled to perfect fluids \cite{lemos1,germano1,gil0,gil1,gil2,pedram}. 
It became clear that, with the exception of the 
radiative perfect fluid, for all other types of fluids the resulting superhamiltonians
have terms that will lead to factor ordering ambiguities when they are written as operators.
In Ref. \cite{nivaldo}, a coordinate transformation was introduced for the models described 
above, such that, in the new 
coordinates, the superhamiltonian had no terms that could lead to factor ordering ambiguities 
at the quantum level. Later, in Ref. \cite{gil}, it was explicitly demonstrated that this coordinate 
transformation is canonical with the aid of the symplectic method. Unfortunately, the transformation
introduced in Ref. \cite{nivaldo} cannot be applied to the important case of a stiff matter 
perfect fluid. The stiff matter perfect fluid has an equation of state of the form, $p = \alpha w$,
with $\alpha = 1$, where $w$ and $p$ are, respectively, the fluid energy density and pressure. This
perfect fluid can also be described by a massless free scalar field.

The great importance of cosmological models where the matter content is represented by a stiff
matter perfect fluid was recognized since its introduction by Zeldovich \cite{zeldovich}. In order
to understand better the importance of this perfect fluid for cosmology, one has to compute its energy
density. In the temporal gauge ($N(t) = 1$), this quantity is proportional to $1/a(t)^6$, where $N(t)$ is 
the lapse function and $a(t)$ is the scale factor. On the other hand, in the same gauge, the energy
density of a radiative perfect fluid is proportional to $1/a(t)^4$. This result indicates that there
may have existed a phase earlier than that of radiation, in our Universe, which was dominated by 
stiff matter. Due to that importance, many physicists have started to consider the implications of 
the presence of a stiff matter perfect fluid in FRW cosmological models. The first important 
implication of the presence of stiff matter in FRW cosmological models is in the relic abundance of
particle species produced after the `Big Bang' due to the expansion and cooling of our Universe 
\cite{turner,joyce,salati,pallis,gomez,pallis1}. The presence of stiff matter in FRW cosmological 
models may also help explaining the baryon asymmetry and the density perturbations of the
right amplitude for the large scale structure formation in our Universe \cite{zeldovich1,joyce1}.
It may also play an important role in the spectrum of relic gravity waves created during inflation 
\cite{sahni}. Since there may have existed a phase earlier than that of radiation which was dominated by 
stiff matter some physicists considered quantum cosmological models with this kind of matter 
\cite{nelson,nelson1}.

In the present work, we extend the works of Refs. \cite{nivaldo} and \cite{gil} to the important 
case of a stiff matter perfect fluid. In order to do that, we consider Friedmann-Robertson-Walker models 
in the presence of a stiff matter perfect fluid and a cosmological constant. The models differ from each 
other by the positive, negative and null curvatures of the spatial sections. We write the superhamiltonian 
of these models using the Schutz's variational formalism \cite{schutz}. We notice that the resulting 
superhamiltonians have terms that will lead to factor ordering ambiguities when they are written as 
operators. In order to remove these ambiguities, we introduce appropriate coordinate transformations and 
prove that these transformations are canonical using the symplectic method \cite{JBW}.

The Friedmann-Robertson-Walker cosmological models are characterized by the
scale factor $a(t)$ and have the following line element,
\begin{equation}  
\label{1}
ds^2 = - N^2(t) dt^2 + a^2(t)\left( \frac{dr^2}{1 - kr^2} + r^2 d\Omega^2
\right)\, ,
\end{equation}
where $d\Omega^2$ is the line element of the two-dimensional sphere with
unitary radius, $N(t)$ is the lapse function, $k=0,\pm 1$ gives the spatial section 
curvature and we are using the natural unit system, where $\hbar=c=8\pi G=1$. The 
matter content of the model is represented by a perfect fluid with four-velocity 
$U^\mu = \delta^{\mu}_0$ in the comoving coordinate system used, plus a cosmological
constant ($\Lambda$). The total energy-momentum tensor is given by,
\begin{equation}
T_{\mu,\, \nu} = (w+p)U_{\mu}U_{\nu} - p g_{\mu,\, \nu} - \Lambda
g_{\mu,\, \nu}\, ,  
\label{2}
\end{equation}
As mentioned above, here, we assume that $p = w$, which is the equation of 
state for stiff matter. 

Einstein's equations for the metric (\ref{1}) and the energy momentum 
tensor (\ref{2}) are equivalent to the Hamilton equations generated by 
a total Hamiltonian, namely,
\begin{equation}
\label{2000}
{\cal H} = -\frac{{P_a}^2\cdot a^2}{12} - 3ka^{4} + 
\Lambda a^{6}+ P_T.
\end{equation}
The variables $P_a$ and $P_T$ are the momenta canonically conjugated to the variables $a$ and $T$. 
Furthermore, the first term in the total Hamiltonian (\ref{2000}) will pose an operator ordering 
ambiguity problem, when it is written as an operator. This ambiguity will affect the quantum energy 
spectrum of the model. The total Hamiltonian (\ref{2000}) is related to the following Lagrangian,
\begin{equation}
\label{2010}
{\cal L}= P_a\dot a + P_T\dot T - V,
\end{equation}
where $V=N\Omega$ is the symplectic potential, with
\begin{equation}
\label{2020}
\Omega = -\frac{{P_a}^2a}{12} - 3ka^3 + 
\Lambda a^5+ \frac{P_T}{a}.
\end{equation}

In order to eliminate the factor ordering ambiguity problem, the following transformation will be used,
\begin{equation}
\label{2030}
a = e^x,\,\,\, p_a = p_x / a,\,\,\, 
\end{equation}
Due to this, the Lagrangian density 
(\ref{2010}) becomes,
\begin{equation}
\label{2040}
{\cal L} = P_x\dot x + P_T\dot T - V,
\end{equation}
where the symplectic potential is $V=N\Omega$, with,
\begin{equation}
\label{2050}
\Omega = -\frac{{P_x}^2}{12}\cdot e^x - 3ke^{3x} + \Lambda e^{5x} + P_T \cdot e^{-x}.
\end{equation}
The symplectic variables are $\xi_i=(x,P_x,T,P_T,N)$. Following the symplectic method in Ref. \cite{JBW} 
and the procedure developed in Ref. \cite{gil}, it is possible to show that the model has a symmetry, 
which is fixed with the introduction of the following gauge fixing term,
\begin{equation}
\Sigma=N - e^x. 
\end{equation}
Due to this, the original first-order Lagrangian (\ref{2040}) is rewritten as,
\begin{equation}
\label{2100}
{\cal L} = P_x\dot x + P_T\dot T + \Sigma\dot\eta - V,
\end{equation}
Now, the symplectic variables are $\xi_i=(x,P_x,T,P_T,N,\eta)$. The corresponding symplectic matrix becomes nonsingular and, from the inverse of the symplectic matrix, the non null Dirac brackets are obtained,
\begin{equation}
\label{2110}
\left\{x,P_x\right\}=\left\{T,P_T\right\}=1.
\end{equation}
The symplectic potential, identified as being the Hamiltonian, is given by,
\begin{equation}
\label{2120}
{\cal H} =-\frac{{P_x}^2}{12} - V_{eff.} + {P_T},
\end{equation}
with,
\begin{equation}
\label{b00280}
V_{eff.}= 3 k e^{4x} - \Lambda e^{6x}.
\end{equation}
At this point, it is important to notice that the Hamiltonian above will produce a Schr\"{o}dinger 
like equation without operator ordering ambiguities, when it is written as an operator. Further, the 
Dirac brackets are equal to the Poisson brackets, allowing us to conclude that the variable 
transformation (\ref{2030}) is a canonical transformation.

\begin{acknowledgements}
E. V. Corr\^{e}a Silva, G. A. Monerat, G. Oliveira-Neto (Researchers of CNPq, 
Brazil), C. Neves and L. G. Ferreira Filho thank CNPq and FAPERJ for partial financial support.
\end{acknowledgements}

\end{document}